\documentclass[preprint,12pt]{elsarticle}

\usepackage{amssymb}
\usepackage{amsmath}
\usepackage{graphicx}
\usepackage{dcolumn}
\usepackage{siunitx} 
\usepackage{bm}
\usepackage{multirow}
\usepackage{ulem}
\usepackage{color}
\usepackage{longtable}
\usepackage{subfigure}
\usepackage{amssymb}
\usepackage{amsmath}
\usepackage{upgreek}
\usepackage{latexsym,epsfig}
\usepackage{float}
\usepackage[figuresright]{rotating}
\usepackage{upgreek}
\usepackage{ulem}
\usepackage{booktabs}
\usepackage{physics}
\usepackage{gensymb}

\journal{Spectrochimica Acta Part B: Atomic Spectroscopy}

\begin{document}

\begin{frontmatter}



\title{Spectral line-shape in collinear laser spectroscopy after atomic charge exchange}

\author[frib,msu]{Adam Dockery\corref{cor1}}\ead{dockery@frib.msu.edu}
\author[frib,msu]{Kei Minamisono}\ead{minamiso@frib.msu.edu}
\author[frib]{Alejandro Ortiz-Cortes}
\author[frib,msu]{Brooke Rickey}\ead{rickey@frib.msu.edu}
\cortext[cor1]{Corresponding author}
\affiliation[frib]{organization={Facility for Rare Isotope Beams},
            addressline={640 S Shaw Lane}, 
            city={East Lansing},
            postcode={48824}, 
            state={MI},
            country={USA}}
\affiliation[msu]{organization={Department of Physics and Astronomy, Michigan State University},
            addressline={567 Wilson Road}, 
            city={East Lansing},
            postcode={48824}, 
            state={MI},
            country={USA}}
\begin{abstract}
Collinear laser spectroscopy experiments on fast, neutral beams have been extensively used for studies on short-lived radioactive nuclei, taking advantage of its high sensitivity. The resulting resonance line-shape is known to show significant distortion, due to the energy exchange during the charge-exchange neutralization process, which can cause large systematic uncertainty in the determined centroid. A model for the line shape was constructed and simulated to be compared to measured Al, Si, and Ni hyperfine spectra. It is shown that the distortion is caused mainly by the transfer of electron into many different energy levels in the projectile atom and subsequent decays, rather than secondary inelastic collisions, which were often assumed in the line shape analysis before. The model can also be applied to other projectile-alkali pairs, providing a reliable line-shape with less fitting parameters than conventional phenomenological models.
\end{abstract}

\begin{graphicalabstract}
\includegraphics[scale=0.4]{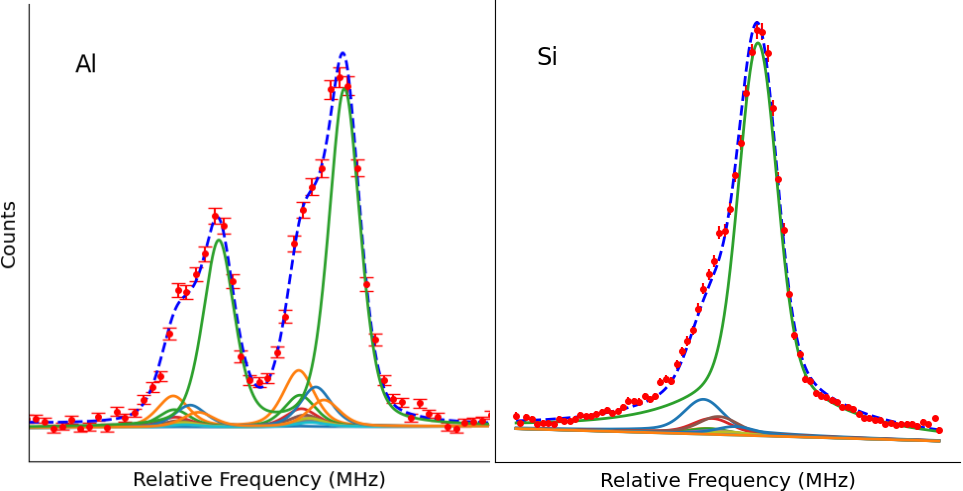}
\end{graphicalabstract}

\begin{highlights}
\item Simulating spectral line-shape of fast beams after atomic charge exchange reactions
\item The simulated line-shape reproduces distortion known to appear in measured spectra
\item Additional free parameters in the conventional fitting model are no longer required
\item The simulated model can be applied to any charge exchange systems
\item Provides a new tool to design laser spectroscopy measurements on fast beams
\end{highlights}


\end{frontmatter}



\section{Introduction}
Collinear laser spectroscopy (CLS) has been used almost exclusively in nuclear structure studies of short-lived radioactive isotopes, taking advantage of the technique's high sensitivity and high resolution \cite{yang23}. The nuclear spins, magnetic dipole moments, electric quadrupole moments, and mean-squared charge radii of radioactive nuclei can be determined, providing some of the highest precision tests of modern nuclear theories \cite{miller19}, and insights into the dense nuclear matter equation of state \cite{pineda21}. Furthermore, the determination of absolute transition frequencies, as well as $A$ and $B$ hyperfine coupling constants, forms a critical test for atomic theories \cite{dockery23}.

In the CLS technique, a fast ion or atom beam at an energy of 10’s of keV is collinearly overlapped with laser light for spectroscopy. Typically, laser resonant fluorescence or optically ionized ions are detected as the resonance signal. Optical transitions in a neutral atom are often used in these measurements rather than transitions in an ionic system, because of the accessible laser wavelengths. 

Atomic charge exchange cells (CEC) at CLS facilities have been developed to create fast, neutral beams via collisions with alkali vapor \cite{klose12}. The development of neutral beams has enabled the laser spectroscopy techniques to be applied to a greater number of elements and their radioactive isotopes. However, the study of neutral beams has introduced a complex resonance line-shape after charge exchange reactions, due to the electron transfer from the alkali into most of the atomic states in the ion to be neutralized, resulting in different velocity classes according to the energy exchange. Therefore, the impact can be significant especially for elements which have sparse electronic level density because the different velocity classes can become clearly separated. 

A significant distortion of the line-shape is often observed in neutral beam spectra \cite{koenig24, sommer22}. The choice of peak profile to account for the distorted line-shape can introduce large systematic uncertainties in the fitted centroid frequencies. Complicated profiles to reproduce the distortion can also lead to increased statistical uncertainty on fitted parameters from the difficulty of the regression with many free fitting parameters. The analysis of these complicated profiles up to now has involved the introduction of phenomenological satellite peaks or skewed profiles to represent secondary inelastic collisions in the CEC \cite{koenig24,sommer22}. This assumption is well-justified for the symmetric charge exchange, e.g. Na beam on sodium vapor. However, it is not evident for the asymmetric charge exchange where the beam and alkali are different elements. As a matter of fact, recent laser spectroscopy measurements of Al following Al on sodium charge exchange reactions have shown line-shape distortion that cannot be adequately accounted for with the satellite peak or skewed line-shapes. Here, we revisit the spectral line-shape after asymmetric charge exchange reactions, and a model is developed that allows hyperfine spectra to be analyzed without relying on free satellite peaks or additional skew parameters. 



\section{Charge Exchange Reactions}
Charge exchange reactions are studied in different velocity classes, where different approaches are necessary. Velocity classes for the ion beam are estimated relative to the Bohr velocity of the exchanged electron in the target atom, which is roughly $10^5-10^6$ m/s for typically used alkali vapors \cite{bransden90}. Low energy reactions are those where the velocity of the beam is much less than the Bohr velocity, and a full wave-mechanical treatment becomes necessary \cite{rapp62}. A discussion of these calculations can be found in \cite{bransden90,massey33,bates58,menshikov82}. High energy reactions are those where the beam velocity is much greater than the Bohr velocity and relativistic effects become important. A variety of perturbation theory and coupled channel approximations are used in this regime \cite{bransden90,macek82,dewangan85}. The charge exchange reactions at beam energies typically used in CLS experiments are considered to be an intermediate region with an ion energy of 10s of keV, corresponding to an ion velocity near the Bohr velocity. These have been mainly studied by semi-classical perturbation theory techniques \cite{bransden90,rapp62,dewangan73,sinha76}. In the intermediate velocity region, charge exchange reactions are typically broken into two classes: the symmetric resonant exchange and the asymmetric non-resonant exchange shown in [Eq. \ref{symm_res_exchange}] and [Eq. \ref{asymm_nonres_exchange}], respectively.
\begin{equation}
\underline{A^{+}} + A \longrightarrow \underline{A} + A^{+}
\label{symm_res_exchange}
\end{equation}
\begin{equation}
\underline{A^{+}} + B \longrightarrow \underline{A} + B^{+} + \Delta E
\label{asymm_nonres_exchange}
\end{equation}
The underlined species in the above equations represents the projectile. For the non-resonant exchange reactions, $\Delta E$ is the difference between the electron entry point energy and a state in $A$. Symmetric exchange reactions have been studied extensively in theoretical models \cite{bransden90,sinha76}. Experiments on the symmetric reactions show overall agreement with the theoretical cross-sections.

\begin{figure}
    \centering
    \includegraphics[scale=0.8]{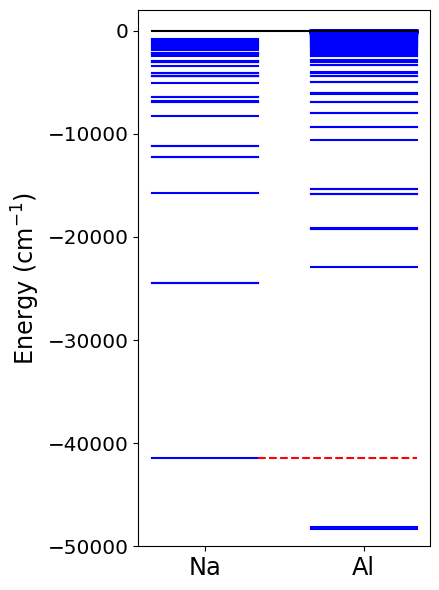}
    \caption{Na and Al Energy Level Matchup. The match-up of Na and Al energy levels relative to their respective ionization potentials. The electron entry point is taken as the Na ground state and is shown in the red dashed line.}
    \label{na_al_matchup}
\end{figure}

Modern CLS experiments typically entail a wide range of projectiles and an alkali vapor, due to the nature of nuclear structure studies to explore many different elements and their isotopes. The asymmetric charge exchange process is almost always a non-resonant exchange process and has been treated theoretically in \cite{rapp62,dewangan73}. As an example, the energy-level matchup of Al and Na for the Al on sodium charge exchange is shown in Fig. \ref{na_al_matchup}. The model from \cite{rapp62,dewangan73} has shown reasonable agreement with total cross sections and relative state populations in experimental works \cite{ryder15,vernon19}. 
In addition to the resonant and non-resonant exchange reactions, secondary inelastic collisions can occur in the interaction between the projectile and alkali vapor. Target alkali excitation processes [Eq. \ref{a+_exc_b}, \ref{a_exc_b}] as well as projectile excitation processes [Eq. \ref{b_exc_a+}, \ref{b_exc_a}] are possible \cite{bendali86, springeramo}.
\begin{equation}
\underline{A^{+}} + B \longrightarrow \underline{A^{+}} - \Delta E_{B \rightarrow B^*}  + B^* \longrightarrow \underline{A^{+}} - \Delta E_{B \rightarrow B^*}  + B + h\nu
\label{a+_exc_b}
\end{equation}
\begin{equation}
\underline{A} + B \longrightarrow \underline{A} - \Delta E_{B \rightarrow B^*}  + B^* \longrightarrow \underline{A} - \Delta E_{B \rightarrow B^*}  + B + h\nu
\label{a_exc_b}
\end{equation}
\begin{equation}
\underline{A^{+}} + B \longrightarrow \underline{A^{+*}} - \Delta E_{A^+ \rightarrow A^{+*}}  + B \longrightarrow \underline{A^{+}} - \Delta E_{A^+ \rightarrow A^{+*}}  + B + h\nu
\label{b_exc_a+}
\end{equation}
\begin{equation}
\underline{A} + B \longrightarrow \underline{A^*} - \Delta E_{A \rightarrow A^{*}}  + B \longrightarrow \underline{A} - \Delta E_{A \rightarrow A^{*}}   + B + h\nu
\label{b_exc_a}
\end{equation}
In the symmetric charge exchange with A = B, for example Na\textsuperscript{+}-Na reaction, processes according to [Eq. 3, 4, 6] are degenerate and cannot be independently identified. The large separation of the ground state energy level in Na\textsuperscript{+} suggests the process in [Eq. 5] would be suppressed, as $\Delta E_{A^+ \rightarrow A^{*+}}$ is a minimum of approximately 33 eV. Additionally, the large energy gap between the first excited doublet and additional excited states in Na suggests only excitations to the first doublet are likely. In this case, a Poisson distribution [Eq. \ref{poisson_distr}] can be used to account for multiple secondary collisions of the degenerate processes, and this model successfully reproduced experimental data in a previous study \cite{bendali86}.
\begin{equation}
    P(n) = \frac{x^n}{n!} e^{-x}
\label{poisson_distr}
\end{equation}
Here, $n$ is the number of secondary collisions and x is a parameter that characterizes the interaction length and alkali vapor density. The asymmetric charge exchange reaction presents a more complicated case, where the reactions in [Eq. \ref{a+_exc_b}, \ref{a_exc_b}] are no longer degenerate with those in [Eq. \ref{b_exc_a}]. In addition, the possibility for many exchange states in the projectile $A$ means reactions in [Eq. \ref{a_exc_b}, \ref{b_exc_a}] must be evaluated for every state in $A$ having a significant exchange cross section.

\section{Experiment}
\begin{figure}
    \centering
    \includegraphics[scale=0.36]{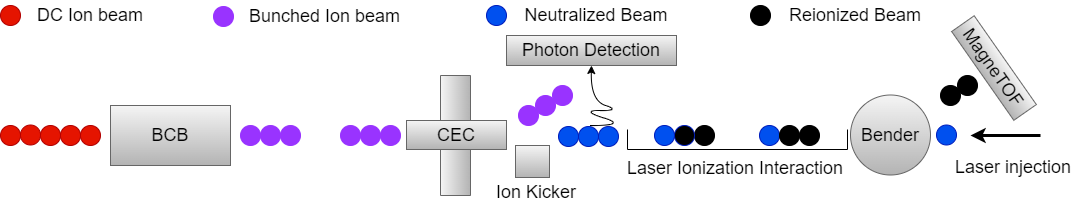}
    \caption{Schematic of the Experimental Beamline at BECOLA. Ions are generated as a DC beam, and then restructured into ion bunches in the BCB. Following the CEC, the neutralized component is isolated and overlapped with laser light in the photon detection (fluorescence) or laser ionization interaction (resonant ionization) regions with corresponding detectors.}
    \label{beamline_schematic}
\end{figure}

CLS measurements were performed at the BEan COoler and LAser spectroscopy (BECOLA) facility at the Facility for Rare Isotope Beams (FRIB), which is shown schematically in Fig. \ref{beamline_schematic} \cite{minamisono13}. Singly-charged ions of the element of interest were generated by a Penning Ionization Gauge (PIG) ion source for offline beam production \cite{ryder15}. The ion beam then entered the radio-frequency beam-cooler-buncher (BCB), where the ions were cooled by collisions with He buffer gas \cite{barquest17}. The ion beam was extracted from the BCB and accelerated to approximately 30 keV by an electrostatic potential. The acceleration compresses the velocity spread in the beam, and therefore, reduces the Doppler broadening of the observed spectra, known as the velocity bunching \cite{kaufman76}.
To measure spectra in the atom, the ion beam was neutralized by collisions with Na vapor in the CEC \cite{klose12}. A small potential was applied to the CEC in order to scan the observed laser frequency in the atom rest frame due to the Doppler shift. The atom beam then entered the laser interaction region where resonant ionization or fluorescence measurements were performed. Details of each measurement type are briefly given below.

\subsection{Fluorescence}
For the fluorescence measurements, the ion beam was either allowed to fly through the BCB without trapping (DC mode) or extracted as a bunched beam (bunched mode). After neutralization, the atom beam overlapped with co- or counter-propagating resonant laser light. Fundamental laser light was produced by a tunable continuous-wave (CW) Ti:Sapphire ring laser (Sirah Matisse), which was locked to the readout of a wavemeter (HighFinesse WSU-30). The wavemeter was calibrated every minute against a frequency stabilized He-Ne laser (SIOS Meßtechnik GmbH). Typical frequency drift over a day measurement period was 1 MHz. The fundamental light was then frequency doubled (SpectraPhysics Wavetrain) for the spectroscopy. Laser light was transported by an optical fiber and was injected from the collinear or anti-collinear side depending on the measurement geometry. The fluorescent light was detected by photomultiplier tubes (PMTs) in the detection region. One detection system contained an ellipsoidal reflector to yield spacial separation between resonant and background photons and improve signal-to-noise ratio \cite{minamisono13}. The other was designed with an oval mirror to increase the detection solid angle of fluorescent photons and minimize background photons \cite{maass20}. This method was used to measure the $3d^94s$ $^3D_3$ $\longrightarrow$ $3d^94p$ $^3P_2^o$ transition in \textsuperscript{58}Ni and the $3s^23p^2$ $^1S_0$ $\longrightarrow$ $3s^23p4s$ $^1P_1^o$ transition in \textsuperscript{28}Si. Details of the measurements are given in \cite{sommer22} and \cite{koenig24}\textcolor{blue}{,} respectively. Additionally, this method was used to measure the $3s^23p$ $^2P_{1/2}^{o}$ $\longrightarrow$ $3s^24s$ $^2S_{1/2}$ transition in \textsuperscript{27}Al which will be detailed elsewhere.

\subsection{Resonant Ionization}
The newly implemented Resonant Ionization Spectroscopy Experiment (RISE) equipment was used to measure the $3s^23p$ $^2P_{1/2}^{o}$ $\longrightarrow$ $3s^25s$ $^2S_{1/2}$ transition in \textsuperscript{27}Al. A brief overview is given here. The atom beam was extracted from the BCB in 100 Hz bunches, and following the CEC, a kicker potential was applied to deflect ions that were not neutralized. Then, the atom beam was overlapped with a co- or counter-propagating resonant laser pulse followed by a counter-propagating non-resonant ionization laser pulse. Resonant laser light was produced by an injection seeded Ti:Sapphire cavity \cite{reponen18}. A single mode Nd:YAG laser (Photonics) with a 10 kHz pulse frequency was used as the pump laser. A tunable CW Ti:Sapphire ring laser (Sirah Matisse) was used as the seed light. The resonant light was always injected from the anti-collinear side, and in the case of a collinear measurement, a 0\degree mirror was used to reflect the light along the same axis. The ionization laser light was produced by a frequency doubled Nd:YAG laser (Merion MW) with a 100 Hz pulse frequency, and the light was injected from the anti-collinear side. The relative timing of the laser pulses and BCB release were controlled by a pulse generator (Quantum Composer). Following the re-ionization, the ion beam passed through a bender to eliminate the neutral component and was measured by an ion detector (MagneTOF ETP Ion Detect). 

\section{Simulation}
Here we develop our simulation of the charge exchange reactions and its resulting effect on spectral line-shape. Throughout, the Al-Na charge exchange reaction and subsequent measurement of the $3s^23p$ $^2P_{1/2}^{o}$ $\longrightarrow$ $3s^25s$ $^2S_{1/2}$ transition in neutral Al is used as an example. The energy levels of Al and Na, relative to each ionization potential (IP), are show in Fig. \ref{na_al_matchup}. Charge exchange reaction simulations are developed in an equivalent form to previous work in sections 4.2 and 4.3 \cite{ryder15}. A more formal linear algebra expression is introduced in section 4.3, so that the simulation can be modified in section 4.4 to account for the kinetic energy spread in the neutral beam. In section 4.5, the kinetic energy spread is developed into a peak profile that can be used to fit data.

\subsection{Assumptions}
The simulation is performed with a series of assumptions to simplify the calculation. The projectile and target are assumed to be in their ground states for the systems considered here. In the case of several low-lying atomic states, a Boltzmann distribution can be used to approximate relative populations of the states. During the charge exchange reaction, it is assumed that the electron can enter into any level in the projectile according to their cross sections. The original work \cite{rapp62} introduces a statistical factor to correct the cross sections by the matching of angular momenta of these states. Due to ambiguity in evaluating this factor, and nonphysical results from attempts to evaluate it, the statistical factor is assumed to be one. The vapor density is assumed to be low such that secondary inelastic collisions have insignificant probability. This is possible by keeping the alkali vapor pressure low and was demonstrated by measurements discussed in section 6.3.

In the development of a simulated model in section 4.5, the broadening of states having excess energy from the reaction is discussed. The excess energy is assumed to be completely absorbed into the projectile due to the large impact parameters typical in these reactions. The additional energy may kick the projectile in any direction, which will change the kinetic energy along the z-axis (which is overlapped with laser light) by a maximum of $\Delta E$ [Eq. \ref{asymm_nonres_exchange}\textcolor{blue}{]}. To approximate the distribution of energies, we assume that $\Delta E$ is the range from tail to center of a normal distribution, which is taken to be $3\sigma$. Thus, the half-width at half-maxima of the broadening is estimated to be $\Delta E /3$.


\subsection{Initial Exchange}
\begin{figure}
    \centering
    \includegraphics[scale=0.45]{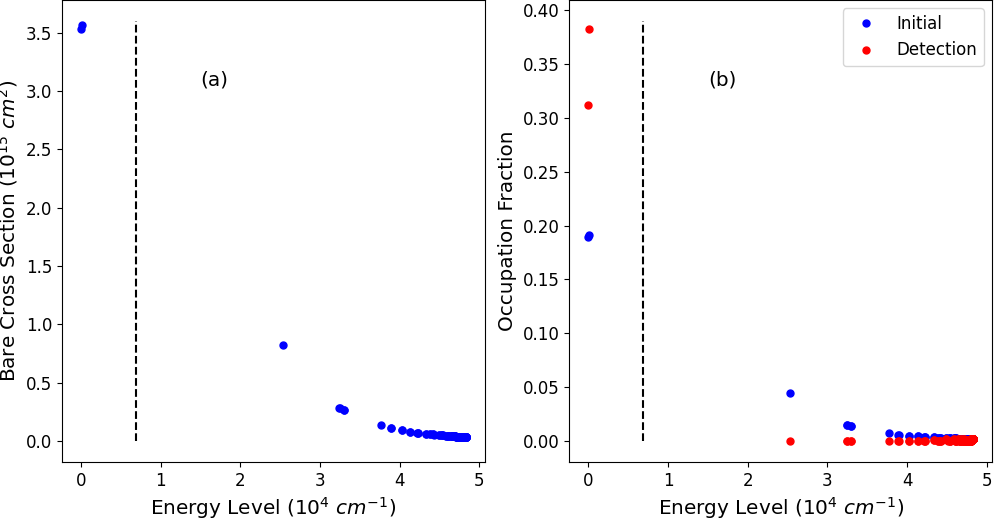}
    \caption{Charge Exchange Cross Sections and Populations for Al-Na Exchange. The non-resonant cross section was calculated for the Al-Na charge exchange reaction with all possible entry states and is shown in (a). The fraction of each state before and after spontaneous decay is shown in (b). The fraction has been normalized so the sum over all states is equal to one.}
    \label{al_cross_sections}
\end{figure}

Charge exchange cross sections in the intermediate velocity region are calculated from a semi-classical perturbation theory following the method of Rapp and Francis with the correction from Dewangan \cite{rapp62, dewangan73}. The cross section equation is shown in [Eq. \ref{cross_section}] 
\begin{equation}
\sigma(b,v) = 2\pi f \int_{0}^{b_{1}}  P(b,v) b db
\label{cross_section}
\end{equation}
Here, \textit{f} is the statistical factor which evaluates the percentage of collisions that result in an allowed transition. The evaluation of \textit{f} is ambiguous and attempts to evaluate it have yielded nonphysical results, so we take \textit{f} to be 1 in this work. The integral is performed over the impact parameter \textit{b}. In the original work \cite{bendali86}, this integral is estimated analytically, but modern computers allow for the integral to be evaluated numerically in short computation times. This integral is evaluated from 0 to infinity in principle, but integrating over too large of \textit{b} values results in nonphysical oscillations \cite{ryder15}. A cutoff value $b_{1}$ is chosen where the integral stabilizes before oscillation. \textit{P(b,v)} is the probability as a function of impact parameter and velocity, which is given in [Eq. \ref{prob_dist}]. 
\begin{equation}
P(b,v) = \sin^2 \left( \sqrt{\frac{8 \pi b^3}{\gamma a_{0}}} \frac{E_i}{\hbar v} \left(1+\frac{a_0}{\gamma b} \right) e^{-\gamma b / a_{0}}\right) \times \sech^2 \left(\sqrt{\frac{a_{0} \pi b}{2\gamma}}\frac{\omega}{v}\right)
\label{prob_dist}
\end{equation}
In the above equation, $\gamma = \sqrt{\frac{E_{i}}{13.6}}$ and $E_{i}$ is some mean between the ionization potentials of the projectile and target; this ambiguity in evaluating $E_{i}$ is present in the original theory \cite{rapp62}. In this work, $E_{i}$ is chosen as the mean between the ionization potential of the projectile and target from the electronic levels they occupy. The alkali target is assumed to be in the ground state, and the projectile entry level is allowed to vary. $a_{0}$ is the bohr radius, $v$ is the ion beam velocity, and $\omega$ is the difference between the energy level of the projectile and target divided by $\hbar$, which are taken relative to their respective ionization potentials. As an example, the Al-Na charge exchange reaction is considered, and the cross section for each Al energy level is evaluated and shown in Fig. \ref{al_cross_sections}(a). Energy levels were taken from the Kurucz database \cite{kurusz}.

\subsection{Decay}
Following the initial exchange, there is a flight distance before the ions meet with laser lights. Therefore, the decay of excited states needs to be modeled to get an accurate population distribution at the measurement location. Redistribution of populations typically stabilizes after a few centimeters flight time, and here, 40 cm is chosen as the flight distance which is the distance for fluorescence measurements at BECOLA. Other longer distances were tested and consistent results were achieved. The column vector $\vec{p}_{t}$, with elements $p_{t,k}$ for the k\textsuperscript{th} energy level, is introduced to represent the population distribution at time t. Immediately after the charge exchange reaction $\vec{p}_{0}$ is defined in [Eq. 10].
\begin{equation}
    \vec{p}_{0} = \sigma_{k} \hat{e}_{k} / \sum \sigma_{k}
\end{equation}
\label{p0_def}
In the above equation, $\hat{e_{k}}$ represents the k\textsuperscript{th} energy level in the projectile, and $\sigma_k$ is the cross section for the electron to enter that same level. The population is normalized to sum to one. Einstein coefficients are taken from the Kurucz database as decay rates \cite{kurusz}. The matrix $D$ represents the spontaneous emission decay processes. As spontaneous emission can only go from higher energy levels to lower, this is an upper triangular matrix (with 0 on the diagonal) with elements given by the Einstein coefficient for the corresponding transition $A_{ij}$ multiplied by the time step $dt$ [Eq. \ref{d_def}].
\begin{equation}
    D_{ij} = A_{ij} dt
\label{d_def}
\end{equation}
The interval $dt$ is chosen such that the change of population for any time step is small, that is $A_{ij}dt << 1$. By introducing conservation of probability, the diagonal elements are given by [Eq. \ref{prob_cons}].
\begin{equation}
    D_{jj} = 1 - \sum_{i=j+1} D_{ij}
\label{prob_cons}
\end{equation}
For a time $t$ after the charge exchange reaction, $\vec{p}_{t}$ is calculated by [Eq. \ref{pt_def}] where $n$ is the number of time steps defined in [Eq. \ref{n_def}].
\begin{equation}
    \vec{p}_{t} = D^n \vec{p}_{0}
\label{pt_def}
\end{equation}
\begin{equation}
    n = \frac{d}{v\times dt}
\label{n_def}
\end{equation}
In the above equation, $d$ is the flight distance and $v$ is the ion velocity. This method was applied to the Al-Na charge exchange reaction, and the population after a 40 cm time of flight is shown in Fig. \ref{al_cross_sections}(b). Lower energy levels tend to increase in population due to the feeding from the higher levels, which in turn decrease in population. It is noted that due to the high level density at higher excitation energy, even though these high lying levels have small initial population, the feeding to the low-lying states becomes significant.

\subsection{Decay with Kinetic Energy States}
\begin{figure}
    \centering
    \includegraphics[scale=0.8]{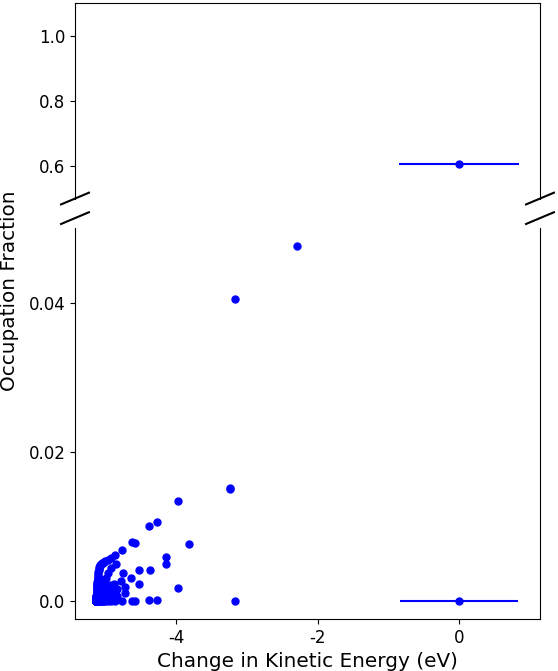}
    \caption{Kinetic Energy Components Following Al-Na Exchange. The Al population was evolved over the 40cm flight path, and the normalized population of each kinetic energy state in the $3s^23p$ $^2P_0^0$ ground electronic state is plotted. The occupation fraction is normalized to sum to one. Kinetic energy states corresponding to $\Delta E > 0$ are shown with x-uncertainty, since the excess energy can kick the projectile in any direction. A point is placed at the middle to represent the average value.}
    \label{kinetic_energy_components}
\end{figure}

In sections 4.2 and 4.3, the typical simulation of the charge exchange and decay is presented, and results that are consistent with previous works on other systems are found \cite{ryder15}. We now note that because of the difference in energy for each state in Al relative to the electron entry point, the kinetic energy of the projectile is varied for every state [Eq. \ref{asymm_nonres_exchange}]. This is not accounted for in the decay simulation in section 4.3. To account for this kinetic energy difference, the initial population after exchange becomes [Eq. \ref{p0k_def}].
\begin{equation}
    p_{0,k} = (\vec{p}_{0} \cdot \hat{e}_{k}) {u}_{k}
\label{p0k_def}
\end{equation}
$\hat{u}_{k}$ denotes the kinetic energy state, and in principle can take on continuous values. However, it is assumed that each projectile ion begins with the same kinetic energy, so $\hat{u}_{k}$ only has one allowed value for each entry energy level $k$. Note that the spontaneous decay matrix $D$ only affects the electron energy level $\hat{e}_{k}$ and not the beam kinetic energy $\hat{u}_{k}$ .The population at a later time $t$ [Eq. \ref{pt_def}] then becomes [Eq. \ref{ptk_def}].
\begin{equation}
    p_{t,k} = \sum_j (D^n)_{kj}(\vec{p}_{0} \cdot \hat{e}_{j}) {u}_{j}
\label{ptk_def}
\end{equation}
This form makes it clear that different kinetic energy states should not be summed into the same $p_{t,k}$, so a new vector $\vec{p'}_{t,k}$ is introduced with elements $p'_{t,k,k'}$ which represents the population of different kinetic energy states $k'$ for a given energy level $k$ at time $t$ [Eq. \ref{ptkk'_def}].
\begin{equation}
    p'_{t,k,k'} = (D^n)_{kk'}(\vec{p}_{0} \cdot \hat{e}_{k'}) {u}_{k'}
\label{ptkk'_def}
\end{equation}
Note that at $t=0$ the value of $p'_{0,k,k'}$ is equal to $p_{0,k}$. Following the above method, $\vec{p'}_{t,k}$ was evaluated for the ground state of Al after the Al-Na charge exchange reaction and the 40 cm flight path. $\vec{p'}_{t,k}$ is re-normalized so that the total population of the ground state is one, and the re-normalized $\vec{p'}_{t,k}$ is plotted in Fig. \ref{kinetic_energy_components}. For reactions with $\Delta E < 0$, the change in energy is assumed to come from the kinetic energy of the ion beam. For reactions with $\Delta E > 0$, it is assumed that the energy is absorbed by the projectile and can kick the projectile in any direction, so a line is drawn to represent the maximum range of energies the neutral beam can have along the z-axis. For the Al-Na exchange, approximately 60\% of the final ground state population exchanged directly to the ground state, while an additional 40\% originated from a higher lying state, and therefore has a different kinetic energy than the direct exchange component.

\subsection{Line-shape Model}
To properly account for the different kinetic energy components in the neutral beam, a modified line-shape profile can be used in regression analysis of measured spectra. CLS measurements are typically analyzed with a pseudo-Voigt profile [Eq. \ref{pseudo-voigt}].
 \begin{equation}
 \begin{split}
     f(x:C,\mu,\sigma_{g},\sigma_{l},\alpha) & = \frac{(1-\alpha)C}{\sigma_{g}} \sqrt{\frac{\text{ln}(2)}{\pi}}\text{exp}\left( {\frac{-(x-\mu)^2 \times \text{ln}(2)}{\sigma_{g}^2}}\right) \\
     & + \frac{\alpha C}{\pi}\frac{\sigma_{l}}{(x-\mu)^2+\sigma_{l}^2}
 \end{split}
 \label{pseudo-voigt}
 \end{equation}
Here, $C$ is the amplitude of the peak, $\mu$ is the centroid of the peak, $\sigma_{g}$ and $\sigma_{l}$ are the Gaussian and Lorentzian half-widths at half-maxima respectively, and $\alpha$ is the Lorentzian fraction. We introduce a separate pseudo-Voigt peak for each kinetic energy state $V_{k'}$, and we fix the $\alpha$ and $\sigma_l$ across all peaks. The amplitude of each $V_{k'}$ is fixed to the normalized value of $p_{t,k,k'}$. For the $\mu$ and $\sigma_g$ of $V_{k'}$, we consider three different cases.
\begin{itemize}
    \item $\Delta E = 0$\textcolor{blue}{:} The beam kinetic energy is unchanged, so a standard pseudo-Voigt peak can be used with $\mu_{k'} = \mu_{\Delta E=0}$ and $\sigma_{g,k'} = \sigma_{g,\Delta E=0}$. This situation only occurs in symmetric resonant charge exchange.

    \item $\Delta E < 0$\textcolor{blue}{:} The additional energy for the charge exchange reaction must come from kinetic energy, and as we assume the alkali vapor to have $v \approx 0$, this energy comes from the projectile kinetic energy. Therefore, the centroid of $V_{k'}$ is shifted by the energy difference $\mu_{k'} = \mu_{\Delta E =0} - \Delta E_{k'}$. The $\sigma_{g,k'}$ is taken to be equal to the standard value $\sigma_{g,k'} = \sigma_{g,\Delta E=0}$.

    \item $\Delta E > 0$: Following the discussion in section 4.1, the excess energy from the charge exchange reaction is assumed to be absorbed by the projectile. The distribution of kinetic energy along the z-axis is assumed to be normal with a half-width at half-maxima of $\Delta E/3$. Therefore, the energy difference is added in quadrature to the $V_{k'}$ Gaussian half-width by $\sigma_{g, k'} = \sqrt{\sigma_{g, \Delta E =0}^2 + \left( \frac{\Delta E_{g, k'}}{3} \right)^2}$. The centroid $\mu_{k'}$ is set equal to the standard value $\mu_{k'} = \mu_{\Delta E=0}$.
\end{itemize}
As CLS spectra are typically analyzed in the beam rest-mass frequency frame, the $\Delta E_{k'}$ can be converted to the rest frequency frame by multiplying by the transition sensitivity [Eq. \ref{transition_sensitivity}] \cite{koenig21}. 
\begin{equation}
\begin{split}
    \frac{\partial \nu_{c/a}}{\partial E_{kin}} & = \frac{2\nu_0}{mc^2} \frac{\nu_{c/a}^2}{\nu_{c/a}^2-\nu_{0}^2} \\ 
    & = \frac{2\nu_{c/a}^3}{mc^2}\frac{\gamma(1 \pm \beta)}{\nu_{c/a}^2\gamma^2(1 \pm \beta)^2 - \nu_{c/a}^2}
\end{split}
\label{transition_sensitivity}
\end{equation}
In the above equation, $\nu_0$ is the rest frame transition frequency which can be separately determined from a collinear-anticollinear measurement \cite{koenig21}, $\nu_{c/a}$ are the collinear and anticollinear resonant laser frequencies, $m$ is the mass of the neutral beam, $\beta$ is the velocity relative to the speed of light, and $\gamma$ is the Lorentz factor. If $\nu_0$ is not known, the bottom form of the equation must be used with $\gamma$ and $\beta$ approximated.

\begin{figure}
    \centering
    \includegraphics[scale=0.45]{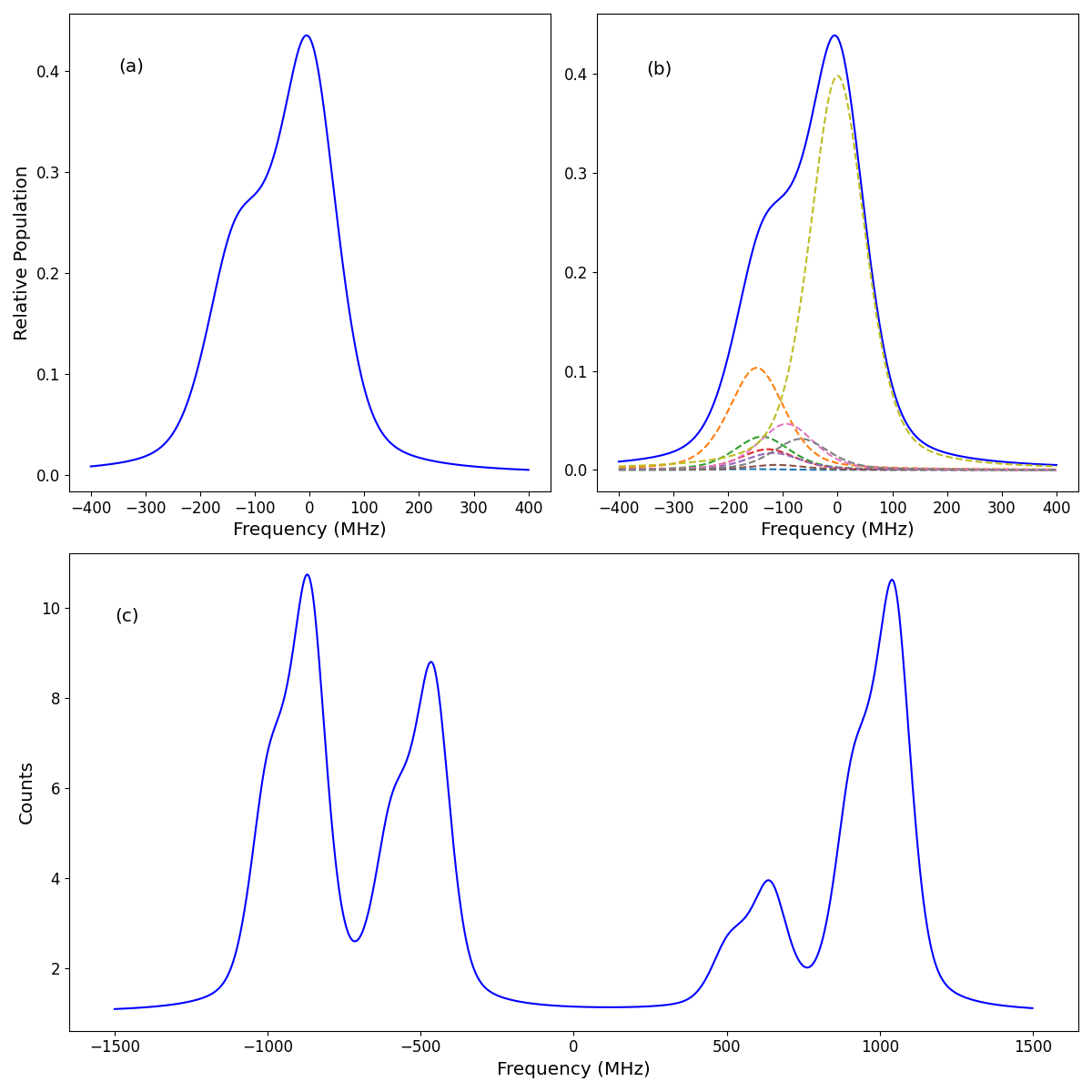}
    \caption{Simulated Spectral Line-Shape for Al. In (a), the peak profile of one of the well-separated hyperfine structure transitions of the $3s^23p$ $^2P_{1/2}^{o}$ $\longrightarrow$ $3s^25s$ $^2S_{1/2}$ transition in Al with 255 different energy components. For comparison, an approximate profile in (b) is shown with components summed into 10 MHz bins (9 components) that is easier to use for fitting. A dotted line is plotted for each of the components in (b). In (c), the simulated hyperfine structure of the $3s^23p$ $^2P_{1/2}^{o}$ $\longrightarrow$ $3s^25s$ $^2S_{1/2}$ transition is shown with the individual peak profiles from (b).}
    \label{al_profile}
\end{figure}

The shape of the peak is strongly affected by the elements in the charge exchange reaction and chosen transition. In Fig. \ref{al_profile}(a), a hyperfine peak for the $3s^23p$ $^2P_{1/2}^{o}$ $\longrightarrow$ $3s^25s$ $^2S_{1/2}$ transition in neutral Al is constructed from the above model following an Al-Na charge exchange reaction. Until now, the typical analysis of such spectra has involved adding a side-peak with the distance and amplitude left as free parameters. Here, the distorted line-shape has been developed from the principles of the reaction without any additional free parameters.
Due to the computational complexity of evaluating a fit with hundreds of peaks, it is practical to reduce the number of peaks used in the fitting function. If the separation between two peaks is sufficiently small compared to the peak widths, the two peaks can be represented by one peak with the amplitude given by the sum of the constituent amplitudes and the centroid given by the center of gravity of the two peaks. In Fig. \ref{al_profile}(b), an approximate peak with 9 components is constructed from summing all components into 10 MHz bins, which is small compared to the typical width of $\approx$ 100 MHz. The difference between fitted parameters determined with these two functions was at most 5 kHz, which is approximately 100 times smaller than the statistical uncertainty achieved in the best measurements of this transition.

In Table \ref{tab:fitvar}, the regression results using different sized bins are compared for a representative run. The regression was performed on a local linux cluster, and computational times vary largely depending on the quality of the data and computer used in the regression. Table \ref{tab:fitvar} shows the computation time as well as the deviation from the full model for the fitted centroid and hyperfine $A$ coupling constants for the ground and excited states. Large deviations only begin with 50 MHz bins (3 components), although the number of required components may vary depending on the transition measured. Following these results, the 10 MHz bins (9 components) were selected, as shown in Fig. \ref{al_profile}(b), since the computation time was significantly reduced without large shifts in any fitted parameters.

\begin{table*}
\centering
\caption{\label{tab:fitvar}Comparison of regression results for a representative run of the $3s^23p$ $^2P_{1/2}^{o}$ $\longrightarrow$ $3s^25s$ $^2S_{1/2}$ transition in \textsuperscript{27}Al. The components have been summed into bins as described in the text. Values for the fitted parameters are given relative to the full model, which is the no bin result. Components with zero amplitude were dropped from the regression.}
\begin{tabular}{cccccc}
 Bin Width&Components&Comp. Time&A $^2P^o_{1/2}$&A $^2S_{1/2}$& Centroid\\
 (MHz)&&(s)&(MHz)&(MHz)&(MHz)\\
 \hline \hline \\

 None&193&277.1&0.000&0.000&0.000\\
 0.1&61&41.6&0.000&0.000&0.000\\
 0.5&36&20.6&0.000&0.000&0.000 \\
 1&26&12.3&0.000&0.000&0.000 \\
 5&13&5.5&0.000&0.000&0.001 \\
 10&9&3.5&0.000&0.000&0.005 \\
 25&5&2.2&0.002&0.004&0.024 \\
 50&3&1.3&0.011&0.037&0.190 \\
 100&2&1.0&0.016&0.162&0.713 \\
 
 \end{tabular}
\end{table*}

\section{Results}

\subsection{HFS Analysis}
\label{hfs_analysis}
Hyperfine spectra were obtained for the $3s^23p$ $^2P_{1/2}^{o}$ $\longrightarrow$ $3s^25s$ $^2S_{1/2}$ and $3s^23p$ $^2P_{1/2}^{o}$ $\longrightarrow$ $3s^24s$ $^2S_{1/2}$ transitions in \textsuperscript{27}Al. The location of each hyperfine peak was fixed by the energy shift equation [Eq. \ref{hyp_energy_shift}].
\begin{equation}
    \Delta E_{\text{hyp}} = - \frac{K_g}{2} A_{g} + \frac{K_e}{2} A_{e}
\label{hyp_energy_shift}
\end{equation}
$A_g$ and $A_e$ are the hyperfine coupling constants for the ground and excited states respectively, and $K = F(F+1) - I(I+1) - J(J+1)$ is evaluated for both the ground and excited states. Since $J = \frac{1}{2}$ for every state in these transitions, both $B$ hyperfine coupling constants are zero. Three different peak profiles were used to analyze the \textsuperscript{27}Al spectra. First, a line-shape was generated according to the method of section 4.5. This analysis method will  be referred to as "simulation."

Next, a main pseudo-Voigt peak [Eq. \ref{pseudo-voigt}] was accompanied by one pseudo-Voigt satellite peak with a free distance and amplitude relative to the first peak. This distance and amplitude were shared among the satellite peaks for each main peak. The satellite peak and main peak shared the same $\alpha$ and $\sigma_{g,l}$. One satellite peak was chosen as the best option by evaluating different numbers of satellite peaks, fixed by the model of the Poisson distribution, with the results of a high precision scan obtained in a separate measurement. This analysis method will be referred to as "one satellite."

Depending on the combination of the projectile and alkali, the asymmetry may also appear more smooth than as a notable satellite peak. In this case, it is common to use a skewed-Voigt profile [Eq. \ref{skewed_voigt}, \ref{z_def}]. 
\begin{equation}
    f(x; C, \mu, \sigma, \gamma, \zeta) = C \frac{\text{Re}[e^{-z^2} \times \text{erfc}(-iz)]}{\sigma \sqrt{2\pi}}\left(1 + \text{erf}[\frac{\zeta(x-\mu)}{\sigma\sqrt{2}}]\right)
\label{skewed_voigt}
\end{equation}
\begin{equation}
    z = \frac{x-\mu+i\gamma}{\sigma\sqrt{2}}
\label{z_def}
\end{equation}
In the above equations, $C$ is the amplitude, $\mu$ is the centroid, $\sigma$ is the Gaussian width, $\gamma$ is the Lorentzian width, and $\zeta$ controls the size of the skew. erfc is the complimentary error function, erf is the error function, and $i$ is the imaginary unit. This analysis method will be referred to as "skewed." 

\subsection{Transition Frequencies and Hyperfine Coupling Constants}
\begin{figure}
    \centering
    \includegraphics[scale=0.28]{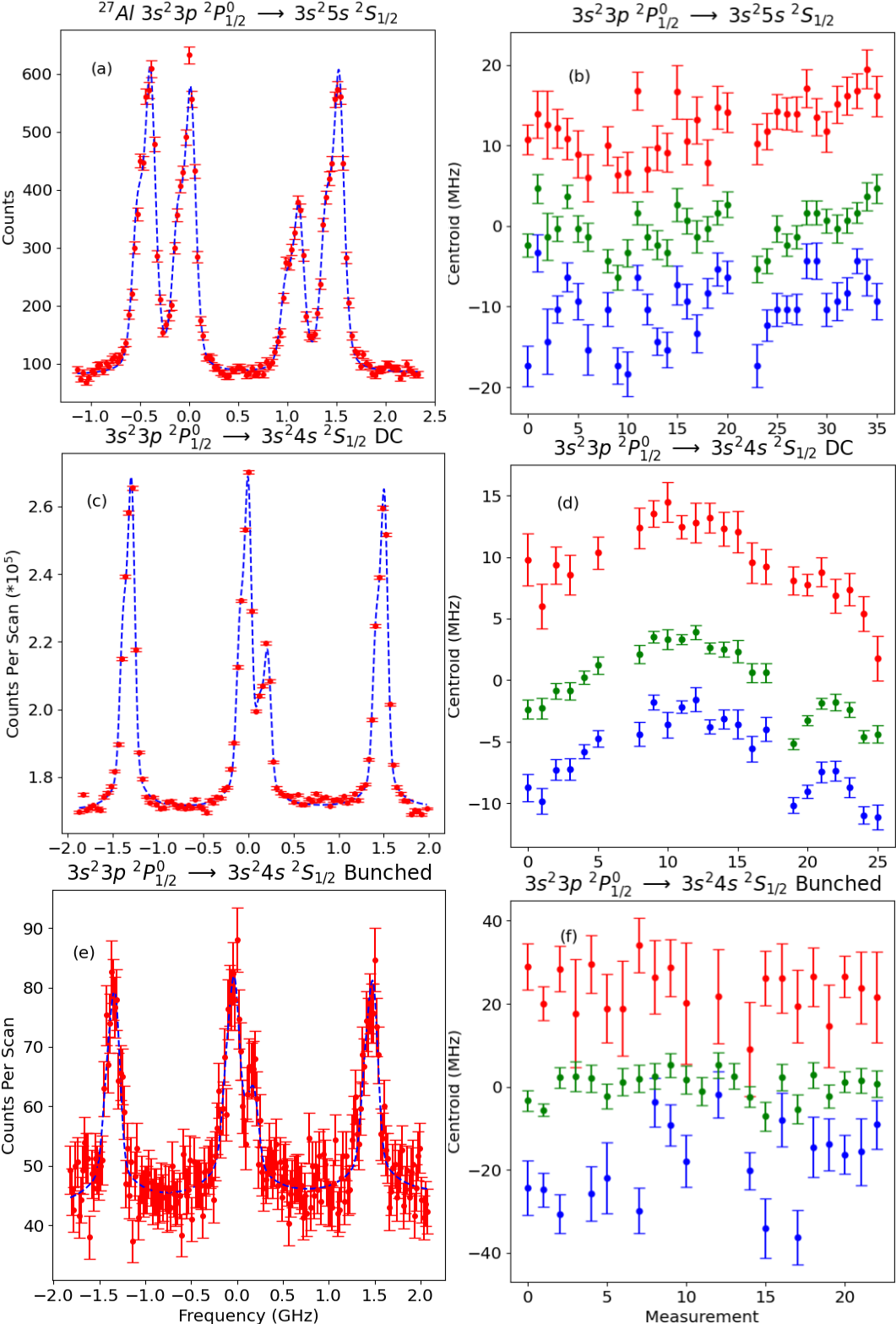}
    \caption{Example Fitted Spectrum and Centroid Scatter in Al. In (a), an example spectrum of the $3s^23p$ $^2P_{1/2}^{o}$ $\longrightarrow$ $3s^25s$ $^2S_{1/2}$ transition is shown, and the determined transition frequencies over 35 independent measurements are plotted in (b) for each analysis method with skewed (red), simulation (green), and one satellite (blue). An arbitrary offset is applied to each analysis set. Similarly, (c) and (e) show example spectrum for the $3s^23p$ $^2P_{1/2}^{o}$ $\longrightarrow$ $3s^24s$ $^2S_{1/2}$ transition in the DC and bunched beam measurement respectively. The corresponding transition frequencies over approximately 25 independent measurements are shown in (d) and (f).}
    \label{centroids_and_spectra}
\end{figure}

Measured hyperfine spectra were evaluated with the above analysis methods to determine the transition frequencies and couplings constants. A representative spectrum and the obtained transition frequencies for the $3s^23p$ $^2P_{1/2}^{o}$ $\longrightarrow$ $3s^25s$ $^2S_{1/2}$ transition in \textsuperscript{27}Al are plotted in Fig. \ref{centroids_and_spectra}(a) and (b) respectively. The mean, weighted standard deviation, and mean fit uncertainty for the transition frequency and coupling constants, found for each analysis method, are given in Table \ref{tab:ris}. The fit uncertainty was scaled by the square-root of the chi-squared goodness of fit per degree of freedom $\sqrt{\chi^2_{dof}}$. The mean fit uncertainty is the average error bar for each measurement in the series. Example spectra for the $3s^23p$ $^2P_{1/2}^{o}$ $\longrightarrow$ $3s^24s$ $^2S_{1/2}$ transition are shown in Fig. \ref{centroids_and_spectra}(c) and (e) for the DC and bunched beam measurement respectively. The determined transition frequencies are shown in Fig. \ref{centroids_and_spectra}(d) and (f), and results are given in Table \ref{tab:fluorescence}. Standard deviation was not evaluated for the DC case as a drift was observed in the resonant frequency throughout the measurements Fig. \ref{centroids_and_spectra}(d).

\begin{table*}
\centering
\caption{\label{tab:ris}Results from the analysis of the $3s^23p$ $^2P_{1/2}^{o}$ $\longrightarrow$ $3s^25s$ $^2S_{1/2}$ transition. All values in the table are given in MHz. Only the last four digits of the centroid values are given, and all mean centroid values are given relative to $1.129898  \times 10^9$ MHz. }
\begin{tabular}{ccccc}
 Parameter&Line-shape&Mean&Std. Dev. &Mean Fit 
 Unc.\\
 
 \hline \hline \\
 Centroid&One Satellite&454.7&3.8&2.2 \\
 &Skewed&516.9&3.4&2.6 \\
 &Simulation&464.0&2.7&1.6 \\
 \\
 $A$ $^2P_{1/2}^o$&One Satellite&502.6&1.3&1.1 \\
 &Skewed&502.5&1.4&1.1 \\
 &Simulation&502.6&1.3&1.1 \\
 \\
 $A$ $^2S_{1/2}$&One Satellite&135.7&1.4&1.0 \\
 &Skewed&136.2&1.3&1.0 \\
 &Simulation&135.5&1.4&1.1 \\
 \hline \hline \\
 
 \end{tabular}
\end{table*}

\begin{table*}
\centering
\caption{\label{tab:fluorescence}Results from the analysis of the $3s^23p$ $^2P_{1/2}^{o}$ $\longrightarrow$ $3s^24s$ $^2S_{1/2}$ transition. All values in the table are given in MHz. Only the last four digits of the centroid values are given, and all mean centroid values are given relative to $7.59905 \times 10^8$ MHz. Standard deviation is not evaluated for the DC centroids, as there was a clear drift present Fig. 6(d).}
\begin{tabular}{cccccc}
 Beam Type&Parameter&Line-shape&Mean&Std. Dev.&Mean Fit Unc.\\
 
 \hline \hline \\
 DC&Centroid&One Satellite&878.8&---&0.8 \\
 &&Skewed&919.6&---&1.4 \\
 &&Simulation&894.4&---&0.6 \\
 \\
 &$A$ $^2P_{1/2}^o$&One Satellite&502.4&0.4&0.4 \\
 &&Skewed&502.4&0.4&0.6 \\
 &&Simulation&502.3&0.3&0.4 \\
 \\
 &$A$ $^2S_{1/2}$&One Satellite&431.9&0.5&0.4 \\
 &&Skewed&431.9&0.5&0.6 \\
 &&Simulation&432.0&0.4&0.4 \\
 \\

 Bunched&Centroid&One Satellite&701.1&9.6&6.1 \\
 &&Skewed&744.8&5.1&8.5 \\
 &&Simulation&719.6&3.5&3.0 \\
 \\
 &$A$ $^2P_{1/2}^o$&One Satellite&502.7&1.6&1.8 \\
 &&Skewed&503.0&1.5&1.9 \\
 &&Simulation&502.6&1.6&1.9 \\
 \\
 &$A$ $^2S_{1/2}$&One Satellite&432.1&1.6&1.9 \\
 &&Skewed&431.9&1.7&1.9 \\
 &&Simulation&432.3&1.4&1.9 \\
 \hline \hline \\
 
 \end{tabular}
\end{table*}

\section{Discussion}
\subsection{Aluminum}
In Fig. 6(a), (c), and (e) the plotted fit is obtained from the simulation method, and a good overall agreement with the observed structure is seen. Different transition sensitivities were used for the analysis of the $3s^23p$ $^2P_{1/2}^{o}$ $\longrightarrow$ $3s^25s$ $^2S_{1/2}$ spectra (29.2 MHz/eV) and $3s^23p$ $^2P_{1/2}^{o}$ $\longrightarrow$ $3s^24s$ $^2S_{1/2}$ spectra (20.2 MHz/eV). The overall agreement in all of these cases suggests the simulation method is accurately representing the processes at play in the Al-Na reaction. 

In Table 1 and 2, the transition frequencies and coupling constants as determined by the regression are compared for the three analysis methods. For each of the three cases, data was taken over one day by measuring the same transition many times. The coupling constants are consistent between all analysis methods regardless of the transition or number of counts. A systematic shift in the transition frequency is found between the three analysis methods. This systematic shift cancels during a collinear-anticollinear measurement \cite{koenig21}, so the determination of the absolute rest frame transition frequency is consistent. The shift will also cancel in the determination of isotope shifts, which reinforces the importance of analyzing each isotope with the same line-shape model.

The fit uncertainty of transition frequency and scatter vary between the three analysis methods. This can be seen in Table \ref{tab:ris} and \ref{tab:fluorescence} by the standard deviation and mean fit uncertainty. In Fig. \ref{centroids_and_spectra}(b),(d), and (f), the trend of each measurement of transition frequency also highlights this. For the $3s^23p$ $^2P_{1/2}^{o}$ $\longrightarrow$ $3s^25s$ $^2S_{1/2}$ transition, the simulation method performs slightly better than the one satellite and skewed line-shape methods. The observed trend throughout the day is similar between all three analyses. 

In the high statistics (DC) case of the $3s^23p$ $^2P_{1/2}^{o}$ $\longrightarrow$ $3s^24s$ $^2S_{1/2}$ transition, all three fit methods perform similarly, with the skewed line-shape method having a slightly larger uncertainty than the other methods. However, in the low statistics (bunched) case, the simulation method performs significantly better than the other analysis methods. This is due to the difficulty in fitting the additional parameters that define the shape of the one satellite and skewed profiles at low statistics. The removal of reliance on free parameters that define the shape is particularly useful in the analysis of spectra obtained from rare isotopes where the count rate is usually low.

\subsection{Silicon, Nickel}
To further evaluate the performance of the simulation, the line-shape for the $3d^94s^3$ $^3D_3$ $\longrightarrow$ $3d^94s^24p$ $^3P_2$ transition in \textsuperscript{58}Ni and $3s^23p^2$ $^1S_{0}$ $\longrightarrow$ $3s^23p4s$ $^1P^o_1$ transition in \textsuperscript{28}Si were simulated. These transitions were measured in previous fluorescence experiments at BECOLA \cite{koenig24,sommer22}.

\begin{figure}
    \centering
    \includegraphics[scale=0.53]{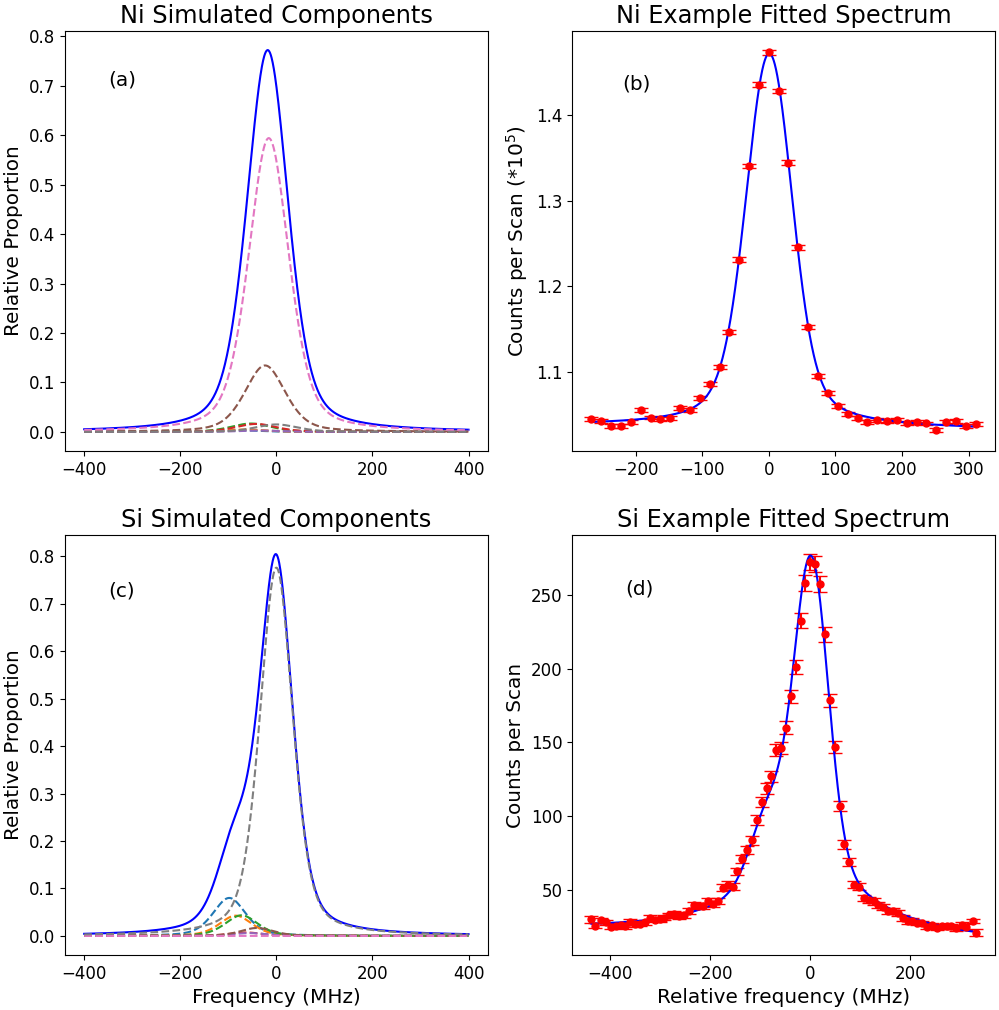}
    \caption{Simulated Line-shape and Example Spectrum for Ni and Si. The simulated line-shape of the $3d^94s^3$ $^3D_3$ $\longrightarrow$ $3d^94s^24p$ $^3P_2$ transition in \textsuperscript{58}Ni is shown in (a) with components summed into 10 MHz bins (8 components), and it is fit to an example spectrum in (b). In (c), the $3s^23p^2$ $^1S_{0}$ $\longrightarrow$ $3s^23p4s$ $^1P^o_1$ transition in \textsuperscript{28}Si is simulated with components summed into 10 MHz bins (8 components) and fit to an example spectrum in (d).}
    \label{si_ni_figure}
\end{figure}

The simulated line-shape for the $3d^94s^3$ $^3D_3$ $\longrightarrow$ $3d^94s^24p$ $^3P_2$ transition in Ni is shown in Fig. \ref{si_ni_figure}(a), and this simulation is fit to a representative measured spectrum in Fig. \ref{si_ni_figure}(b), where the fitted function does not contain parameters for the line-shape other than the overall amplitude. The model correctly predicts a nearly symmetric line-shape with a slight skew to the low frequency side. The small separation of components is reflective of the small energy differences between atomic states in Ni and the small transition sensitivity (14.8 MHz/eV) for this transition. The transition sensitivity is smaller than that for Al because of the heavier mass and longer laser wavelength used (352 nm). In the previous work \cite{sommer22}, an exponential tail was used to account for the skew in the peak.

In Fig. \ref{si_ni_figure}(c), the simulated line-shape is shown for the $3s^23p^2$ $^1S_{0}$ $\longrightarrow$ $3s^23p4s$ $^1P^o_1$ transition in Si, having a transition sensitivity of 19.5 MHz/eV. Small side-peak components are shifted by 50 MHz to 150 MHz depending on the energy difference from the main peak to the low frequency side. This causes a characteristic kink on the low frequency side of the peak, which matches measured experimental data. A representative spectrum is fit to experimental data with the simulation model in Fig. \ref{si_ni_figure}(d) without parameters for the line-shape but the overall amplitude. In the previous work \cite{koenig24}, a Poisson model with four side peaks was used to account for the shape. 

The agreement of the simulation model with measured spectra in Al, Ni, and Si gives further confidence to the applicability of the present model. Each peak shows a different form of distortion: a strong shoulder in Al (one side peak), a slight skew in Ni (exponential tail), and a subtle kink in Si (four side peaks from a Poisson distribution). All of these shapes are reproduced with the same simulated model in this work. This suggests that the simulated model can be widely applied to different CLS measurements using a variety of projectiles and exchange medium.

Additionally, the strong distortion observed in Al and Si suggests that light-mass elements are more strongly affected than mid-mass or heavy elements. There are two likely causes for the large distortion at lighter masses. The sparse level density means population-contributing atomic states are more clearly separated in energy, and often, there is not a near resonant state to transfer electrons to. Additionally, since the transition sensitivity is inversely related to mass, light mass elements have a more clear separation of contributing states in the frequency space used for analysis.

\subsection{Vapor Pressure Variation}
At BECOLA, the pressure of Na vapor is controlled by adjusting the current applied to the heater on the Na reservoir. The configuration of the CEC means the exact vapor pressure of the Na is not known and does not reach a saturated state \cite{klose12}. However, the set current can be used as a pseudo for the vapor pressure. The $3s^23p$ $^2P_{1/2}^{o}$ $\longrightarrow$ $3s^25s$ $^2S_{1/2}$ transition in \textsuperscript{27}Al was used to study the effect of the Na vapor pressure on the detected line shape. The current was raised from a low density state (heater at 2.8 A) to a high density state (heater at 3.9 A) in 0.1 A steps. Measurements made while the reservoir was heating are labeled in 0.05 A increments. The heater was then lowered back to the low density state in equivalent steps to check the consistency of the measurement. 

\begin{figure}
    \centering
    \includegraphics[scale=0.65]{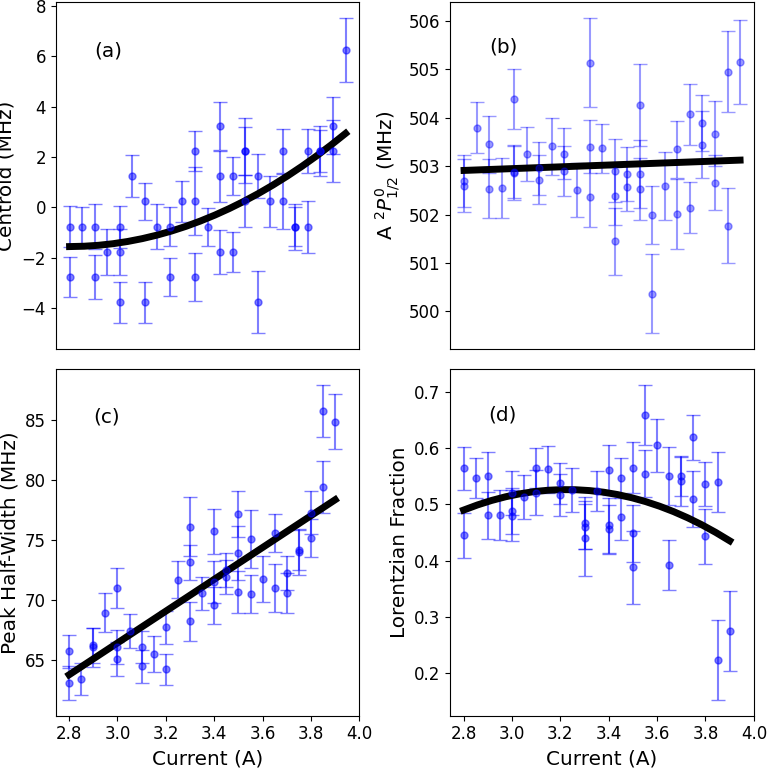}
    \caption{CEC Temperature Dependence of Fitting Parameters. The hyperfine structure centroid (a), $A$ hyperfine parameter for the $^2P_{1/2}^0$ state (b), the peak half-width (c), and Lorentzian fraction (d) are shown for the $3s^23p$ $^2P_{1/2}^{o}$ $\longrightarrow$ $3s^25s$ $^2S_{1/2}$ transition in \textsuperscript{27}Al at various Na reservoir heating currents. The solid line in each plots is the best fit of data to show the trend. The current is used as a pseudo tracker for the Na vapor density in the beamline.}
    \label{cec_temp_figure}
\end{figure}

The measured hyperfine spectra were analyzed using the simulation method as defined in section \ref{hfs_analysis}. The Gaussian and Lorentzian half-widths were constrained to be equal to reduce the frequency of nonphysical fits on the width parameters\textit{}. For several runs, the width constraint was tested against fits with no constraint and consistent results were found. Fig. \ref{cec_temp_figure} includes the fitted results of key parameters for the measurements at various currents. The centroid, shown in Fig. \ref{cec_temp_figure}(a), stays constant until large currents where a shift to higher frequency can be seen. A similar shift at high currents is seen for the Lorentzian fraction in Fig. \ref{cec_temp_figure}(d). The $A$ hyperfine parameter for the ground state does not show any significant trend as seen in Fig. \ref{cec_temp_figure}(b), and a similar trend was seen in the $A$ hyperfine parameter for the excited state which is not plotted here. The peak half-width, Fig. \ref{cec_temp_figure}(c), steadily increases as the current is increased, which suggests that the number of secondary inelastic collisions [Eq. \ref{a+_exc_b}-\ref{b_exc_a}] is increasing and the model is treating it as a broadening. A best fit line is included for each plot only to illustrate the overall trend.

These results suggest that the simulation method is a robust analysis tool provided the CEC temperature is kept below the point where large shifts in the centroid are seen. The shift in the centroid and Lorentzian fraction can be interpreted as the point where secondary collisions become significant enough to be important and deviate from the line-shape obtained by the simulated method. A full analysis in the high vapor pressure regime must include the secondary collisions. As discussed in section 2, an asymmetric exchange reaction with elements A-B has a large number of secondary collisions possible leading to many varying excitation energies. Therefore, a simple Poisson distribution cannot be used to represent the line shape as used in the symmetric charge exchange. A theoretical model for the secondary collision cross sections is highly desired in order to include these in the simulated line-shape, enabling more seamless analysis over the wide range of alkali vapor pressure.

\subsection{Line-shape Tuning}
As shown above, the observed spectral line-shape following the Al-Na exchange is complicated, and cannot be fully represented with one satellite peak or a simple skew. For a simpler analysis, the exchange alkali vapor may be modified to give a more favorable line-shape. This may be particularly useful for the case of spectra obtained from rare isotopes, especially if the CEC is operated at a vapor density with significant secondary collisions. 

\begin{figure}
    \centering
    \includegraphics[scale=0.52]{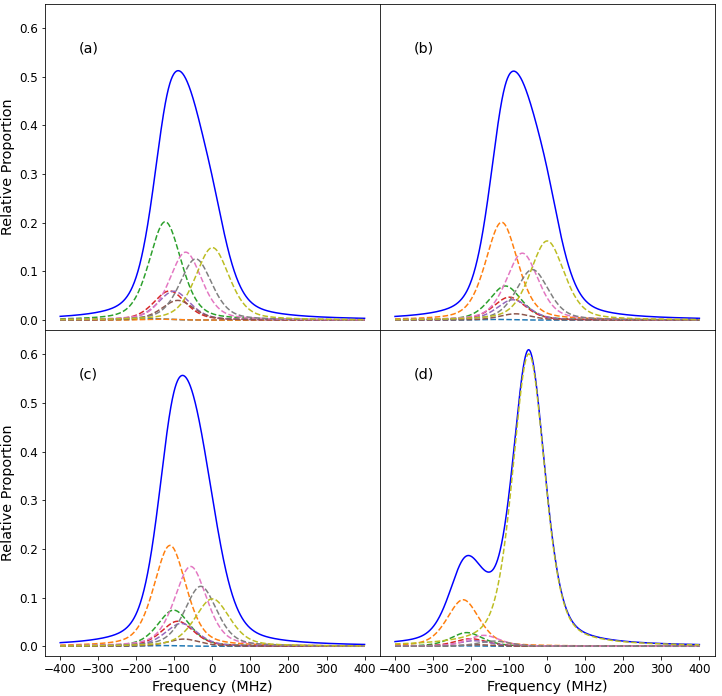}
    \caption{Simulated Line-Shape for Al with Other Exchange Partners. The hyperfine peak line-shape for the $3s^23p$ $^2P_{1/2}^{o}$ $\longrightarrow$ $3s^25s$ $^2S_{1/2}$ transition was simulated for different possible charge exchange vapors equivalent to Fig. 5(b). Simulations were performed with alkali vapor (a) potassium (b) rubidium (c) cesium, and (d) magnesium.}
    \label{al_lineshapes}
\end{figure}

In Fig. \ref{al_lineshapes}, the line-shape of the $3s^23p$ $^2P_{1/2}^{o}$ $\longrightarrow$ $3s^25s$ $^2S_{1/2}$ transition in \textsuperscript{27}Al is modeled for different vapors in the CEC. A more smooth skew is observed if Na is replaced with an alkali metal having a smaller ionization potential like K, Rb, or Cs, shown in Fig. \ref{al_lineshapes}(a), (b), and (c) respectively. If instead an alkali earth metal like Mg is used, Fig. 9(d), a clear separation between the satellite peaks and main peak is observed since Mg has a larger ionization potential than Na. This shape would be more easily fit by the one-satellite peak method, but the fit becomes more complicated with more complex hyperfine spectrum. In the design of laser spectroscopy measurements with a neutral beam, the selection of the type of alkali vapor to optimize the line-shape may be a valuable consideration. This is another significant application of the simulations developed in this work. 

\section{Conclusion}
The asymmetric atomic charge exchange process was simulated to model the distorted line-shape often observed in CLS measurements. The distortion before was conventionally accounted for phenomenologically by a Poisson distribution or skewed peak profile according to the analysis on the symmetric charge exchange, in spite of their different mechanisms for the distortion. By taking into account the various kinetic energy components of the projectile after the transfer of electron from the alkali to many levels in the projectile atom in charge exchange reactions, a simulated resonance line-shape was constructed. The simulated line-shape reproduces the line shape of measured Al, Si, and Ni hyperfine spectra well, and performs better in determining the centroid of resonance than the conventional analysis methods often considering only secondary inelastic collisions.

The simulated line-shape performs particularly better in fitting hyperfine spectra with low statistics, where experimental line-shape is not well resolved, because the simulated line shape removes the reliance on additional free parameters to determine its shape.  This is nearly always the case in the study of rare isotopes far from stability, which are difficult to produce with their small production cross sections and short lifetimes.

The present model can be applied to charge exchange reactions involving a variety of projectiles and targets. The method developed here will be a valuable tool in the analysis of hyperfine spectra as well as in designing laser spectroscopy measurements for easy to analyze line-shape, especially for those of short-lived rare isotopes.

The simulation method, however, shows deviation at high vapor densities where the role of secondary inelastic collisions becomes significant. We note that in the case of isotope shift measurements the deviation will cancel as long as both isotopes are measured at the same alkali vapor density and resulting spectra are fitted by the same line-shape profile. A full treatment at these vapor densities must include processes of secondary inelastic collisions, for which theoretical cross sections are required, which are currently not available. Updated theoretical cross sections for the charge exchange processes including the secondary collisions are highly desired to enable seamless analysis of line shape over wide range of charge exchange vapor pressure.


\section{Acknowledgements}
This work is supported in part by the National Science Foundation Grant No. PHY-21-11185, and DOE Office of Science Award Number DE-SC0000661. The authors would like to thank Joshua Nicholson and Michael Gajdosik for help expressing the equations in section 4.3 and 4.4.

\clearpage

\bibliographystyle{elsarticle-num} 
\bibliography{sources.bib}

\providecommand{\noopsort}[1]{}\providecommand{\singleletter}[1]{#1}%
\begin{thebibliography}{10}
\expandafter\ifx\csname url\endcsname\relax
  \def\url#1{\texttt{#1}}\fi
\expandafter\ifx\csname urlprefix\endcsname\relax\def\urlprefix{URL }\fi
\expandafter\ifx\csname href\endcsname\relax
  \def\href#1#2{#2} \def\path#1{#1}\fi

\bibitem{yang23}
X.~F. Yang, S.~J. Wang, S.~G. Wilkins, R.~F.~G. Ruiz, Laser spectroscopy for the study of exotic nuclei, Prog. Part. Nucl. Phys. 129 (2023) 104005.

\bibitem{miller19}
A.~Miller, K.~Minamisono, A.~Klose, D.~Garand, C.~Kujawa, J.~Lantis, Y.~Liu, B.~Maa\text{\ss}, W.~Nazarewicz, W.~N{\"o}rterh{\"a}user, S.~Pineda, P.-G. Reinhard, D.~Rossi, F.~Sommer, C.~Sumithrarachchi, A.~Teigelh{\"o}fer, J.~Watkins, Proton superfluidity and chare radii in proton-rich calcium isotopes, Nature Phys. 15 (2019) 432--436.

\bibitem{pineda21}
S.~Pineda, K.~K{\"o}nig, D.~Rossi, B.~Brown, A.~Incorvati, J.~Lantis, K.~Minamisono, W.~N{\"o}rterh{\"a}user, J.~Piekarewicz, R.~Powel, F.~Sommer, Charge radius of neutron=defiicnet \textsuperscript{54}ni and symmetry energy constraints using the difference in mirror pair charge radii, Phys. Rev. Lett. 127 (2021) 182503.

\bibitem{dockery23}
A.~Dockery, K.~K{\"o}nig, J.~Lantis., Y.~Liu, K.~Minamisono., S.~Pineda, R.~Powel, Hyerfine structure of low-lying triplet states in \textsuperscript{45}sc ii, Phys. Rev. A 108 (2023) 052816.

\bibitem{klose12}
A.~Klose, K.~Minamisono, C.~Geppert, N.~Fr{\"o}mmgen, M.~Hammen, J.~Kr{\"a}mer, A.~Krieger, P.~Mantica, W.~N{\"o}rterh{\"a}user, S.~Vinnikova, Tests of atomic charge-exchange cells for collinear laser spectroscopy, Nucl. Inst. and Meth. A 678 (2012) 114--121.

\bibitem{koenig24}
K.~K{\"o}nig, J.~Berengut, A.~Borschevsky, A.~Brinson, B.~Brown, A.~Dockery, S.~Elhatisari, E.~Eliav, R.~G. Ruiz, J.~Holt, B.~Hu, J.~Karthein, D.~Lee, Y.~Ma, U.~Meißner, K.~Minamisono, A.~Oleynichenko, S.~Pineda, S.~Prosnyak, M.~Reitsma, L.~V. Skripnikov, A.~Vernon, A.~Zaitsevskii, Nuclear charge radii of silicon isotopes, Phy. Rev. Lett. 132 (2024) 162502.

\bibitem{sommer22}
F.~Sommer, K.~K{\"o}nig, D.~Rossi, N.~Everett, D.~Garand, R.~de~Groote, J.~Holt, P.~Imgram, A.~Incorvati, C.~Kalman, A.~Klose, J.~Lantis, Y.~Liu, A.~Miller, K.~Minamisono, T.~Miyagi, W.~Nazarewicz, W.~N{\"o}rtersh{\"a}user, S.~Pineda, R.~Powel, P.~Reinhard, L.~Renth, E.~Romero-Romero, R.~Roth, A.~Schwenk, C.~Sumithrarachchi, A.~Teigelh{\"o}fer, Charge radii of 55,56ni reveal a surprisingly similar behavior at n = 28 in ca and ni isotopes, Phy. Rev. Lett. 129 (2022) 132501.

\bibitem{bransden90}
B.~H. Bransden, Charge exchange in ion-atom collisions, Cont. Phys. 31 (1990) 19--33.

\bibitem{rapp62}
D.~Rapp, W.~Francis, Charge exchange between gaseous ions and atoms, J. Chem. Phys. 37 (1962) 2631--2645.

\bibitem{massey33}
H.~S.~W. Massey, R.~A. Smith, The passage of positive ions through gases, Proc. R. Soc. Lond. A 142 (1933) 142--172.

\bibitem{bates58}
D.~R. Bates, R.~McCarroll, Electron capture in slow collisions, Proc. R. Soc. Lond. A 245 (1958) 175--183.

\bibitem{menshikov82}
L.~I. Men'shikov, Ion-atom charge exchange at low energies, Zh. Eksp. Teor. Fiz. 85 (1983) 1159--1167.

\bibitem{macek82}
J.~Macek, S.~Alston, Theory of electron capture from a hydrogenlike ion by a bare ion, Phy. Rev. A 26 (1982) 250--270.

\bibitem{dewangan85}
D.~P. Dewangan, J.~Eichler, Boundary conditions and the strong potential born approximation for electron capture, J. Phys. B: Atom. Mol. Phys. 18 (1985) L65.

\bibitem{dewangan73}
D.~P. Dewangan, Dependence of symmetrical charge transfer cross section on ionization potential, J. Phys. B: At. Mol. Phys. 6 (1973) L20--L23.

\bibitem{sinha76}
S.~Sinha, J.~N. Bardsley, Symmetric charge transfer in low-energy ion-atom collisions, Phy. Rev. A 14 (1976) 104--112.

\bibitem{ryder15}
C.~A. Ryder, K.~Minamisono, H.~B. Asberry, B.~Isherwood, P.~F. Mantica, A.~Miller, D.~M. Rossi, R.~Strum, Population distribution subsequent to charge exchange of 29.85 kev ni+ on sodium vapor, Spectrochim Acta Part B At. Spectrosc. 113 (2015) 16.

\bibitem{vernon19}
A.~Vernon, J.~Billowes, C.~Binnersley, M.~Bissell, T.~Cocolios, G.~Farooq-Smith, K.~Flanagan, R.~G. Ruiz, W.~Gins, R.~de~Groote, {\'A}.~Kozor{\'u}s, K.~Lynch, G.~Neyens, C.~Ricketts, K.~Wendt, S.~Wilkins, X.~Yang, Simulation of the relative atomic populations of elements 1 $\leq$ z $\leq$ 89 following charge exchange tested with collinear resonance ionization spectroscopy of indium, Spec. Acta Part B 153 (2019) 61--83.

\bibitem{bendali86}
N.~Bendali, H.~Duong, P.~Juncar, J.~S. Jalm, J.~Vialle, Na+-na charge exchange processes studied by collinear laser spectroscopy, J. Phys. B: At. Mol. Phys. 19 (1986) 233--238.

\bibitem{springeramo}
T.~Kirchner, A.~L. Ford, J.~F. Reading, Springer handbook of atomic, molecular, and optical physics 2nd edition, Springer, 2023, Ch. 53 Ion-atom and atom-atom collisions, pp. 785--794.

\bibitem{minamisono13}
K.~Minamisono, P.~Mantica, A.~Klose, S.~Vinnikova, A.~Schneider, B.~Johnson, B.~Barquest, Commisioning of the collinear laser spectroscopy system in the becola facility at nscl, Nucl. Instrum. Methods Phys. Res. A 709 (2013) 85--94.

\bibitem{barquest17}
B.~R. Barqeust, G.~Bollen, P.~F. Mantica, K.~Minamisono, R.~Ringle, S.~Schwarz, C.~Sumithrarachchi, Rfq beam cooler and buncher for collinear laser spectroscopy of rare isotopes, Nucl. Instrum. A 866 (2017) 18--28.

\bibitem{kaufman76}
S.~L. Kaufman, High-resolution laser spectroscopy in fast beams, Optic. Comm. 17 (1976) 309.

\bibitem{maass20}
B.~Maaß, K.~K{\"o}nig, J.~Kr{\"a}mer, A.~Miller, W.~N{\"o}rterh{\"a}user, F.~Sommer, A 4$\pi$ fluorescence detection region for collinear laser spectroscop, arXiv: 2007.02658 (2020).

\bibitem{reponen18}
M.~Reponen, V.~Sonnenschein, T.~Sonoda, H.~Tomita, M.~Oohashi, D.~Matsui, M.~Wada, Towards in-jet resonance ionization spectroscopy: An injection-locked titanium: Sapphire laser system for the palis-facility, Nucl. Instr. and Meth. in Phys. Res. A 908 (2018) 236--243.

\bibitem{kurusz}
R.~Kurusz, B.~Cell, \href{https://lweb.cfa.harvard.edu/amp/ampdata/kurucz23/sekur.html}{1995 atomic line data kurucz (smithsonian astrophysical observatory, cambridge, ma)}.
\newline\urlprefix\url{https://lweb.cfa.harvard.edu/amp/ampdata/kurucz23/sekur.html}

\bibitem{koenig21}
K.~K{\"o}nig, K.~Minamisono, J.~Lantis, S.~Pineda, R.~Powel, Beam energy determination via collienar laser spectroscopy, Phys. Rev. A 103 (2021) 032806.

\end{thebibliography}

\end{document}